\documentstyle[11pt,epsf,paspconf]{article}

\begin{document}

\title{IRAS\,0421+0400: jets crossing an ISM/IGM interface ?}

\author{W. Steffen, A.J. Holloway}
\affil{Department of Physics and Astronomy, University of Manchester,
    Manchester M13 9PL, UK}

\author{A. Pedlar, D.J. Axon\altaffilmark{1}}
\affil{Nuffield Radio Astronomy Laboratories, University of Manchester,
       Jodrell Bank, Macclesfield, Cheshire SK11 9DL, UK}

\altaffiltext{1}{ESA secondment, Space Telescope Science Institute,
                 Baltimore, MD 21218, USA}

\begin{abstract}
The emission lines \index{emission lines} \index{optical emission} in
the active galaxy
\index{active galaxies} IRAS\,0421+0400 \index{IRAS\,0421+0400} show a
dramatic ($\approx$ 900 $\rm km s^{-1}$) increase in the velocity
spread at the position of radio hotspots \index{hotspots} which are
located at the beginning of extended radio lobes \index{radio lobes}.
We study a model which explains this phenomenon as the result of a jet
emerging through the boundary between the interstellar
\index{interstellar medium} and intergalactic medium . A similar
scenario has previously been suggested as an explanation for wide
angle tail radio sources (WAT's) \index{WAT}.  Based on our model, we
simulate the longslit spectra \index{longslit spectra} of these
regions and compare the results with the observations and find that it
can explain most of the details in the observed longslit spectra.
\end{abstract}


\section{Introduction}
The active galaxy IRAS\,0421+0400 was first detected with IRAS
\index{IRAS}(Soifer et al, 1984) and has a redshift of z=0.046
(Beichman et al, 1985; Hill et al, 1988). It shows emission line
dominated spiral structure \index{spiral structure} and a
Seyfert-2-type spectrum \index{Seyfert galaxies}. The radio structure
is very unusual, consisting of a central 1\,arcsec double source in
the central region and large symmetrically bent lobes ($\sim$
25\,kpc).  The position of the hotspots as well as the the orientation
and bending of the lobes (Beichman et al, 1985; Hill et al, 1988;
Holloway et al, 1996; Steffen et al, 1996) appear to be associated
with the underlying optical spiral structure of the galaxy.

Jets in FR\,I radio galaxies \index{FR\,I radio galaxies} can flare
\index{flaring jets}
very abruptly and show very large opening angles up to $90^{\circ}$ in
diffuse lobes or tails (O'Donoghue et al. 1993).  These structures
often bend very near the flaring point as is the case for
IRAS\,0421+0400. Norman et al (1988) and Loken et al (1995) have
modelled this phenomenon in wide angle tail radio galaxies (WAT) in
terms of a supersonic jet passing through a shock in the ambient gas
where the jet flow becomes subsonic. The jet is then disrupted and
entrains external gas, which becomes turbulent and large and small
scale eddies then develop. Such a shock in the ambient medium could be
due to a supersonic galactic wind moving into the surrounding
intergalactic medium.  We suggest that a similar scenario applies to
IRAS\,0421+0400\ at the position of the radio hot spots. Here a mixing
layer \index{mixing layer} develops around the jet in which ambient
gas is accelerated. The optically radiating gas should therefore be
concentrated in a layer forming an opening quasi-conical surface,
possibly with large scale eddies at some point as suggested by
hydrodynamical simulations.

Owen et al (1990) conducted a search for optical line emission from
the flaring regions in WAT sources but found no significant emission
from the 5 objects they studied.  If our interpretation is correct,
then IRAS\,0421+0400\ is an important test object allowing to study
such a transition region in the optical regime, thereby providing
important kinematical \index{kinematics} information from spatially
resolved spectra.

\noindent
\section{The model and optical observations}

\begin{figure}[t]
\label{outfl.fig}
\centering
\mbox{\epsfclipon\epsfysize=6cm\epsfbox[0 0 430 376]{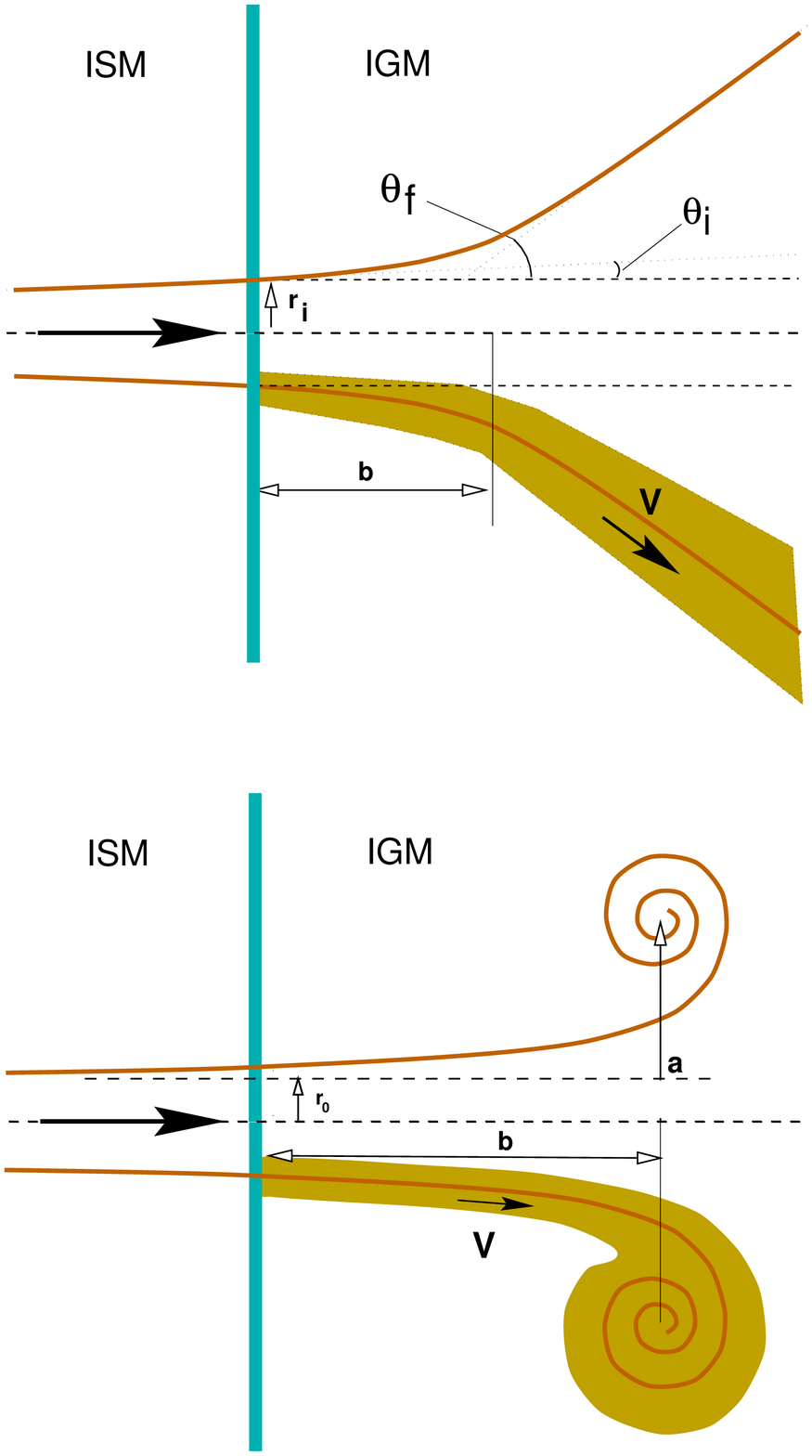}}
\caption{
In this line drawing the geometric model parameters of the flow are
illustrated. They are the initial radius of the circular cross section
$r_i$, the distance $b$ of the eddy from the transition region where
the emission is assume to start, and the distance $a$ from the axis.
}
\end{figure}

We model the longslit emission line spectra using a simple
parameterized description of the emission and velocity field of the
ionized gas flow. In our model we regard the extended emission line
source around the radio hot spots as a collimated outflow which flares
when passing through the boundary between the ISM and the IGM (see
Fig.\,2). The radial shape of the flow line is parameterized
as a gradually opening spiral (to simulate the effect of possible
eddies) given by
\begin{eqnarray}
z_{(s)}  &=& - \frac{a}{s}   \cos(s)
             + \frac{a}{s_0} \cos(s_0) \\
r_{(s)}  &=& (r_0 + a (1-\sin(s)/s))/\epsilon \\
\epsilon &=& (1-e \cos(\phi))/(1-e) .
\end{eqnarray}
Here ($z$,$r$,$\phi$) are cylindrical coordinates, and $s$ is a
normalized position parameter along the spiral (the subscript 0
indicates initial values).  The parameter $a$ is the radial distance
of the centre of the eddy from the axis, while $\epsilon$ allows for
an elliptic cross section of excentricity $e$ as a simple
approximation to non-axisymmetry.

\begin{figure}[t]
\centering
\mbox{\epsfysize=9cm\epsfbox[0 230 540 610]{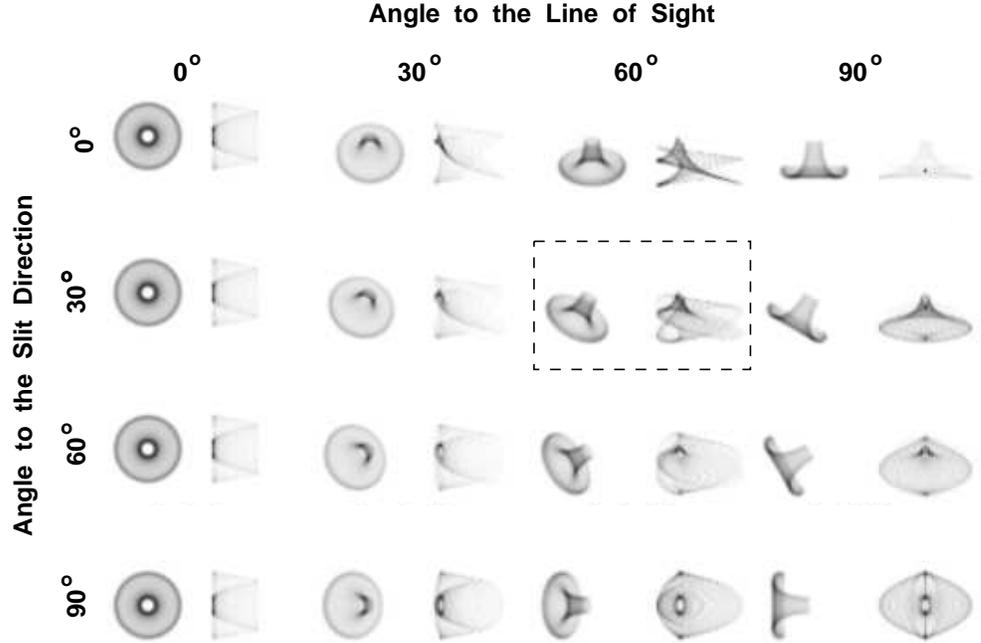}}
\caption{
A series of simulations \index{simulations} of an opening outflow
is shown varying the orientation of the axis with
respect to the line of sight (left to right) and with respect to the
slit orientation. The slit runs from top to bottom and covers the
whole emission. Represented are pairs of images (left) and longslit
spectra (right). The simulation marked with a dashed box is shown with
more details in the next Figure. In this series it is the best
match to the observations at the southern hotspot region in
IRAS\,0421+0400, and was used as a starting point for more detailed
simulations which are compared to the observations below.
}
\label{mosaic.fig}
\end{figure}

\begin{figure}[t]
\centering
\mbox{\epsfysize=5cm\epsfbox[0 75 600 300]{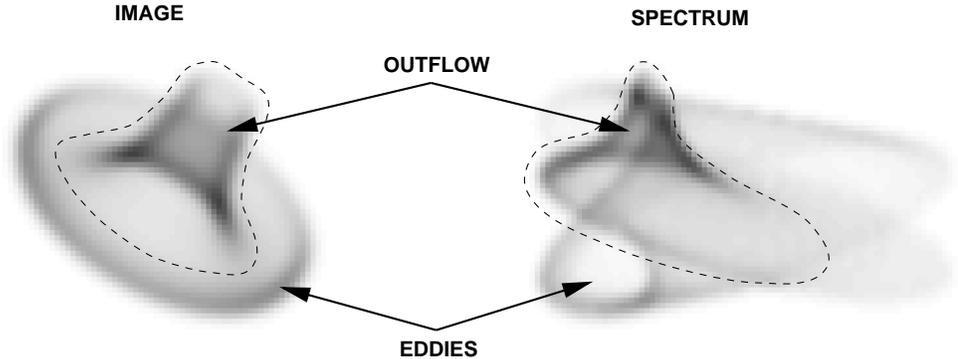}}
\caption{
The simulation marked with a dashed line in the previous Figure
is shown in more detail. The outflowing material roughly marked with
a dashed line in the image simulation (left) basically produces the
correspondingly marked region in the longslit spectrum (right).
Everything outside this region (and some contribution to the inside
region) is originated in the backflow and eddy.
}
\label{detail.fig}
\end{figure}

\begin{figure}[htb]
\centering
\mbox{\epsfysize=15cm\epsfbox[0 0 356 616]{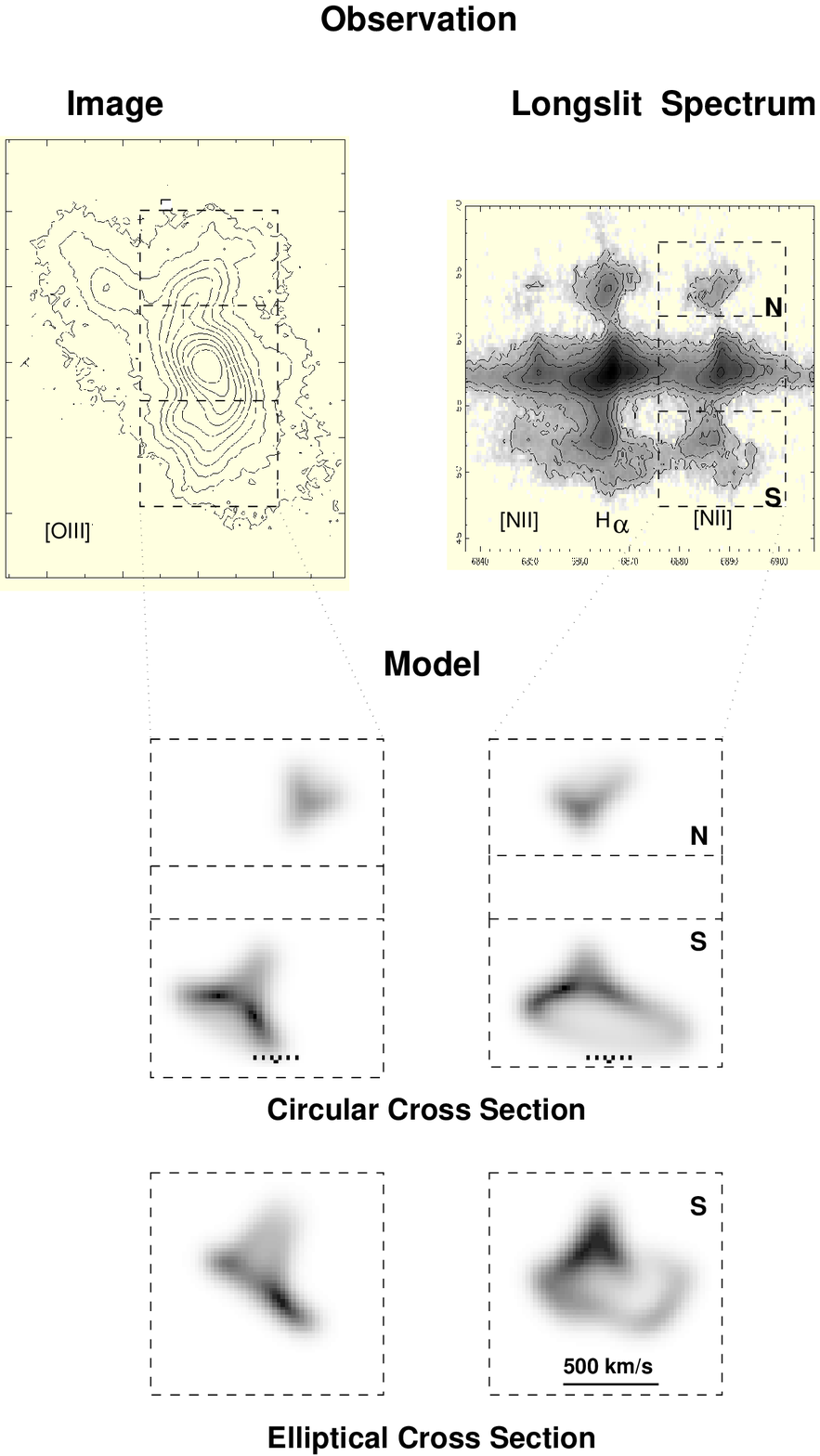}}
\caption{
The top panels show the observed preliminary emission line image
([OIII], left) and the longslit spectra (H$_{\alpha}$-complex, right)
at the same scale. Observe the flaring of the spectra at
approx. 5\,arcsec to the north and south of the centre. The middle
panels show the predicted model image of the flaring region (left) in
the case of an axisymmetric outflow and the corresponding spectrum
(assuming constant velocity along the opening cone of emission line
gas). Note the bifurcations ("snake tongues") in the observed
[OIII]-image which we identify with the opening outflow.  The bottom
row shows a similar simulation of the southern hot spot region, but
with an elliptic cross section. All simulations where smoothed to a
resolution comparable those of the observations.
}
\label{observe.fig}
\end{figure}

The longslit line profile is most dependent on the orientation of the
outflow with respect to the observer's line of sight and the
orientation of the spectrometer slit. The detailed shape of the
outflow and the change in emissivity as a function of distance from
the starting point only influence the detail of the simulated spectrum
and not the gross features.  Therefore, we assume a constant
emissivity and velocity along the flow line and a gaussian
transverse emissivity distribution of FWHM$=0.5 r_0$.

In Figure \ref{mosaic.fig} we show a series of results for
axisymmetric outflows varying the angle with respect to the line of
sight (left to right) and orientation with respect to the direction of
the spectrometre slit (which covers the whole structure). A
logarithmic greyscale is used to emphasize the faint eddies. We find
that the structure of the longslit spectrum depends dramatically on
the orientation of the outflow. For most orientations the opening
section of the outflow produces a bright, asymmetric, and also opening
feature in the spectrum which is often ``V''-shaped. The eddies
generally result in a low brightness, diffuse distribution around the
brighter emission arising from the outflow.

\begin{table}
\caption{The geometric parameters and the velocity $v$ of the simulations
which we compare with the observations are given. The first two rows are
axisymmetric simulations of the northern (N) and southern (S) with
ellipticity $e=0$. ($\Theta$,$\Phi$,$\Psi$) are the Euler angles
giving the orientation in space. The other parameters are as shown
in a previous Figure.}
\begin{center}\scriptsize
\begin{tabular}{llllllll}
\tableline
e   &   & $v$  &   $\Theta$   & $\Phi$        & $\Psi$        & $b/a$  &
$r_0/a$ \\
\tableline
0 & (N)  & 350  & $65^{\circ}$  & $0^{\circ}$   & $30^{\circ}$  & 0.6 & 0.02 \\
0 & (S)  & 500  & $-75^{\circ}$ & $0^{\circ}$   & $30^{\circ}$  & 0.8 & 0.02 \\
0.8 & (S)& 500  & $95^{\circ}$  & $135^{\circ}$ & $215^{\circ}$ & 0.7 & 0.02 \\
\tableline
\end{tabular}
\end{center}
\end{table}

\noindent
The simulation marked with a dashed rectangle compares quite well with
the observed structure in the southern extended emission line region
of IRAS\,0421+0400. However, the observations show no structures
resembling the emission due to the eddies in the simulations (as shown
in more detail in Fig.~\ref{detail.fig}). We conclude that the
emission line gas in IRAS\,0421+0400 is unlikely to be turbulent on
large scales.

In Fig.~\ref{observe.fig} an axisymmetric simulation is shown which
reproduce the observed spectra of the hotspot regions in some detail
(the values of the geometric parameters and the velocity are given in
Table 1). We note, however, that part of the simulated red wing of
the southern region is too bright compared to the blue wing. As shown
in the same figure, an improvement to this situation is found by
introducing an elliptic cross-section and an emissivity cut-off
shortly after the flow turns back.

In Fig.~\ref{observe.fig} we also compare the predicted emission line
image in the hotspot region with the corresponding area in the
[O\,{\sc iii}] 5007-\AA\ image. The latter is a composite of three
images of 30 minutes exposure time each, obtained at the
Anglo-Australian Telescope (AAT) with an
[O\,{\sc iii}] 5007-\AA\ filter (bandwidth 70\AA).  The seeing was
varying between 1 and 2\,arcsec between the individual exposures.  It
confirms the structure observed previously by Beichman et al (1985),
but for the first time shows the predicted ``V''-shape of the northern
hotspot region and there is some indication for a similar, though less
symmetric structure in the south. This supports our interpretation of
an opening outflow of emission line gas in the hotspot regions.

\section{Conclusions}
A simple kinematic model of an opening outflow reproduces the
structure found in the longslit emission line spectrum of the
hotspot regions in IRAS\,0421+0400. The predicted optical image
structure of the same regions is confirmed by recent deep [O\,{\sc
iii}] line-imaging. If the proposed model of a jet crossing a shocked
boundary between the interstellar and intergalactic medium is correct,
then IRAS\,0421+0400 provides a unique possibility to study this
phenomenon at optical wavelengths.

\end{document}